\documentclass[reprint,aps,showkeys,nofootinbib]{revtex4-2}

\usepackage{graphicx}
\usepackage{dcolumn}
\usepackage{bm}
\usepackage{hyperref}
\usepackage{amsmath}
\usepackage{amssymb}
\usepackage{amsfonts}
\usepackage{resizegather}
\usepackage{caption}
\usepackage[british]{babel}

\usepackage{color}
\usepackage{ulem}
\usepackage{cancel}
\usepackage[export]{adjustbox}
\usepackage[newitem,newenum,defblank]{paralist}

\newcommand{\lmatt}{\mathcal{L}_\text{matt}}

\begin{document}

\title{The matter Lagrangian of an ideal fluid.}

\author{Sergio Mendoza}
\email{sergio@astro.unam.mx}
\author{Sarah\'{\i} Silva}
\email{sgarcia@astro.unam.mx}
\affiliation{Instituto de Astronom\'{\i}a, Universidad Nacional Aut\'onoma de M\'exico, AP 70-264, Ciudad de M\'exico 04510, M\'exico}

\date{\today}


\begin{abstract}
  We show that the matter Lagrangian of an ideal fluid equals (up to a
sign --depending on its definition and on the chosen signature of the metric) 
the total energy density of the fluid, i.e. rest energy density plus 
internal energy density. 
\end{abstract}

\keywords{General relativity and gravitation; Fundamental problems and
general formalism; Canonical formalism, Lagrangians and variational
principles}

\maketitle

\section{Introduction}
\label{introduction}

  Since the pioneering work by \citet{taub01} and by Fock \& Kemmer
in their 1955 monograph about gravitation (with English translation
versions in \citet{fock1959} and \citet{fock2015}) it has become
important to calculate the value of the matter Lagrangian used in
relativistic theories of gravity, particularly when dealing with extended
theories of gravity with curvature-matter couplings as first explored
by~\citet{Goenner}, later by~\citet{Allemandi} and~\citet{bertolami08}
and the constructions of~\citet{frlm,harko-lobo-book} deriving in
couplings of the trace of the energy momentum tensor with the curvature
of space-time~\citep{Harko1,barrientos18}.

  The value of the matter Lagrangian has been known to
have values (up to a sign depending on the sign definition
of the matter action and thus the energy-momentum tensor
and the chosen signature of the metric of space-time)
of~\citep[e.g][]{fock2015,hawking-ellis,bertolami09,minazzoli12,harko14}
\( e \) or~\citep[e.g][]{taub02,schutz70,brown93} \( p \) and even the
trace of the energy-momentum tensor~\citep[e.g.][]{ferreira20}, where \(
e \) is the total energy matter density, i.e. mass energy density plus
internal energy density -which for the referred cases were obtained only
for an isentropic or a barotropic fluid, and  \( p \) the pressure of
the fluid .  For the case of the Hilbert action, both values for the
matter Lagrangian are commonly claimed to yield the same Einstein field
equations, but as explained by~\citet{Harko10} the choice \( p = 0 \)
yields a null matter Lagrangian and this appears to be an inconsistency
for extended theories of gravity where non-null matter Lagrangians are
to be selected.  More general studies using multifluids are beginning to
shed some light on the construction of a matter Lagrangian using general
thermodynamical arguments~\citep{gavassino}.

  Using well known modern techniques of calculus of variations on the
definition of the matter Lagrangian, we show in the present work that
for the case of an ideal fluid the matter Lagrangian equals (up to a
sign) the total energy density of the fluid which consists of its rest
mass energy density  plus its internal energy density.
In Section~\ref{action-section} we set up the matter action and define
the energy-momentum tensor as a function of the variations of the matter
Lagrangian.  In Section~\ref{ideal} we write down some basic thermodynamic
relations which will be used in Section~\ref{matter-lagrangian} in order
to obtain the final value of the matter Lagrangian through the use of the
mass continuity equation of general relativistic hydrodynamics.  Finally
in Section~\ref{remarks} we conclude and give examples of the matter
Lagrangian useful for many astrophysical and cosmological applications
useful in extended theories of gravity with curvature-matter couplings.

\section{Action}
\label{action-section}

  The matter action is given by~\citep[see e.g.][]{landau-fields}:

\begin{equation}
  S = \pm \frac{ 1 }{ c } \int{ \lmatt \, \sqrt{-g} \, \mathrm{d}^4 x },
\label{action}
\end{equation}

\noindent where \( \lmatt \) represents the matter Lagrangian.  The
non-positive determinant of the metric \( g_{\alpha\beta} \) is given
by \( g \) and \( c \) is the speed of light.  In the previous equation
we have written down a \( \pm \) sign in the definition of the matter
action \( S_\text{matt} \) since there is no general consensus about
its definition. In what follows we use a signature (\(+,-,-,-\)) for
the metric.  Greek space-time indices vary from \( 0 \) to \( 4 \) and
spatial Latin indices vary from \( 1 \) to \( 3 \).  Einstein's summation
convention is used all over this work.  As it is
traditionally done, we assume that the matter
Lagrangian is a function of the metric tensor \( g^{\alpha\beta} \) only 
and not of its
derivatives \( \partial g^{\alpha\beta} / \partial x^\lambda \).

  The metric, Hilbert or Belifante-Rosenfeld energy-momentum tensor \( T_{\alpha\beta} \) is then defined 
through the variations \( \delta S \) of the matter 
action~\eqref{action} with respect to \( \delta
g^{\alpha\beta} \) and is given by~\citep[see
e.g.][]{landau-fields,harko-lobo-book,franklin17,nuastase19}:

\begin{equation}
  T_{\alpha\beta} =\pm \frac{ 2 }{ \sqrt{-g} }\frac{ 
    \delta \left( \sqrt{-g} \, \lmatt 
    \right) }{ \delta g^{\alpha\beta}}.
\label{energy-momentum}
\end{equation}


%

%
%

  Using the fact that~\citep{landau-fields}:

\begin{equation}
  \delta g = g g^{\alpha\beta} \delta g_{\alpha\beta} = - g g_{\alpha\beta}
  \delta g^{\alpha\beta},
\label{t00}
\end{equation}

\noindent it follows that equation~\eqref{energy-momentum} can be written
as:

\begin{equation}
 T_{\alpha\beta} = \pm 2 \frac{ \delta \lmatt }{ \delta g^{\alpha\beta} } \mp
   g_{\alpha\beta} \lmatt.
\label{em-co}
\end{equation}

\noindent Note that since \( \delta \left( g^{\alpha\beta} g_{\alpha\beta}
\right) = 0 \) then

\begin{equation}
  \delta g_{\alpha\beta} = - g_{\alpha\mu} g_{\beta\nu} \delta
  g^{\mu\nu},
\label{deltag}
\end{equation}

\noindent and with this, equation~\eqref{energy-momentum} can be written
as:

\begin{equation}
 T^{\alpha\beta} = \mp \frac{ 2 }{ \sqrt{-g} }\frac{ 
    \delta \left( \sqrt{-g} \, \lmatt 
    \right) }{ \delta g_{\alpha\beta}} 
\label{energy-momentum-contra}
\end{equation}

\noindent and so\footnote{Since the operator 
\begin{equation}
  \frac{ \delta }{ \delta g^{\alpha\beta} } = 
    \frac{ \delta g^{\mu\nu} }{ \delta g^{\alpha\beta} } 
    \frac{ \partial }{  \partial g^{\mu\nu}  },
\label{foot01}
\end{equation}

\noindent and 

\begin{equation}
  \delta g^{\mu\nu} = \delta \left(  \delta^\mu_\alpha
    \delta^\nu_\beta g^{\alpha\beta} \right) = \delta^\mu_\alpha
    \delta^\nu_\beta \, \delta g^{\alpha\beta},
\label{foot02}
\end{equation}

\noindent relation~\eqref{foot01} can be written as:

\begin{equation}
  \frac{ \delta }{ \delta g^{\alpha\beta} } = 
    \frac{ \partial }{  \partial g^{\alpha\beta} },
\label{foot03}
\end{equation}

\noindent In other words, the \( \delta /
\delta g^{\alpha\beta} \) operator that appears in
equations~\eqref{energy-momentum}-\eqref{em-contra} can be substituted
by \( \partial / \partial g^{\alpha\beta} \).  }:

\begin{equation}
 T^{\alpha\beta} = \mp 2 \frac{ \delta \lmatt }{ 
     \delta g_{\alpha\beta} } + g^{\alpha\beta} \lmatt.
\label{em-contra}
\end{equation}

\section{Ideal fluids}
\label{ideal}

  The appropriate form of the first law of thermodynamics for relativistic
fluids can be written as \citep[see e.g.][]{landau-fluids}:

\begin{equation}
  \mathrm{d}\left( \frac{ e }{ \rho } \right) =T \mathrm{d}\left( \frac{
  \sigma }{ \rho } \right) - p \mathrm{d}\left( \frac{ 1 }{ \rho } \right),
\label{first-law}
\end{equation}

\noindent where the total energy density \( e = \rho c^2 + \xi \)
contains the rest energy density \( \rho c^2 \) and a pure internal
energy density term \( \xi \).  The pressure is represented by \( p \)
and the fluid or gas (baryonic) mass density is \( \rho \).  The entropy
density of the fluid in the previous relation is written as \( \sigma \).

  An ideal fluid is that for which no heat is exchanged between its
components and so, the fluid moves adiabatically~\citep[see
e.g.][]{landau-fluids,acheson,gallavotti,falkovich,tooper}, i.e. \( \mathrm{d}
\left( \sigma / \rho  \right) = 0 \).  In other words, the first law of
thermodynamics for an ideal fluid is given by:

\begin{equation}
  \mathrm{d}\left( \frac{ e }{ \rho } \right) =  - p 
    \mathrm{d}\left( \frac{ 1 }{ \rho } \right),
\label{first-law-tempo}
\end{equation}

\noindent or:

\begin{equation}
  \frac{ \mathrm{d} e }{ e + p } = \frac{ \mathrm{d} \rho }{ \rho }.
\label{first-law-ideal}
\end{equation}

\noindent At this point it is important to mention that the previous two
relations are also valid for an isentropic fluid, i.e. one for which \(
\sigma / \rho = \text{const.} \) This follows from the fact that 
an isentropic fluid is adiabatic.  However, an adiabatic fluid 
is not necessarily isentropic~\citep{landau-fluids}.

  For the case of an ideal fluid, the energy-momentum tensor takes the
form \citep[see e.g.][]{landau-fields,landau-fluids}:

\begin{equation}
  T_{\alpha\beta} = \left( e + p \right) u_\alpha u_\beta - p
  g_{\alpha\beta},
\label{em-fluid}
\end{equation}

\noindent where the four velocity \( u_\alpha \) of the fluid 
satisfies the relation:

\begin{equation}
  u_\alpha u^\alpha = 1.
\label{four-velocity}
\end{equation}

\section{Matter Lagrangian for an ideal fluid}
\label{matter-lagrangian}

  If there are no (baryonic) mass sources, the fluid satisfies a continuity
equation given by~\citep{landau-fluids}:

\begin{equation}
  \nabla_\alpha \left( \rho u^\alpha \right)=0,
\label{continuity}
\end{equation}

\noindent where \( \nabla_\alpha \) is the covariant derivative. The
previous equation can be written as:

\begin{equation}
  \frac{ 1 }{ \sqrt{-g} } \frac{ \partial \left( \sqrt{-g} \, \rho 
    u^\alpha \right) }{ \partial x^\alpha } = 0, 
\label{continuity-tmp01}
\end{equation}

\noindent and so

\begin{equation}
  \delta\left( \sqrt{-g} \, \rho u^\alpha \right) =0. 
\label{continuity-tmp02}
\end{equation}

  In other words:

\begin{equation}
  u^{\alpha}\delta \rho=-\frac{\rho}{\sqrt{-g}} u^{\alpha}\delta
  \sqrt{-g}-\rho \delta u^{\alpha}.
\label{continuity-tmp03}
\end{equation}

  Using the fact that \( \delta \left( g_{\alpha\beta} u^\alpha u^\beta 
\right) = 0 \) and with the help of equation~\eqref{deltag}  it follows
that:

\begin{displaymath}
  2 g_{\alpha\beta} \, \delta u^\beta = - u^\beta \delta g_{\alpha\beta} =
  u^\beta g_{\alpha\mu} g_{\beta\nu} \delta g^{\mu\nu} = u_\nu
  g_{\alpha\beta} \delta g^{\beta\nu},
\end{displaymath}

and so:

\begin{equation}
 2 \delta u^\alpha = u_\beta \delta g^{\alpha\beta}.
\end{equation}

  Substitution of this last result in equation~\eqref{continuity-tmp03} and
with the help of relation~\eqref{t00} yields:

\begin{equation}
  \delta \rho = \frac{ 1 }{ 2 } \rho \left( g_{\alpha\beta} - u_\alpha
    u_\beta \right) \delta g^{\alpha\beta}, 
\label{deltarho}
\end{equation}

\noindent which means that:

\begin{equation}
  \frac{ \delta }{ \delta g^{\alpha\beta} } = \frac{ \delta \rho }{
    \delta g^{\alpha\beta} } \frac{ \mathrm{d} }{ \mathrm{d} \rho } =
    \frac{ 1 }{ 2 } \rho \left( g_{\alpha\beta} - u_\alpha u_\beta \right) 
    \frac{ \mathrm{d} }{ \mathrm{d} \rho },
\label{chain-rule_01}  
\end{equation}

\noindent for an adiabatic fluid.  Using this relation,
equation~\eqref{em-co} --or equivalently relation~\eqref{em-contra}-- can
be written as:

\begin{equation}
  T_{\alpha\beta} = \left( \pm \rho \frac{ \mathrm{d} \lmatt  }{ \mathrm{d} \rho }
    \mp \lmatt \right) g_{\alpha\beta} \mp \rho u_\alpha u_\beta  \frac{
      \mathrm{d} \lmatt  }{ \mathrm{d} \rho },
\label{em-final}
\end{equation}

\noindent for an ideal fluid.

  From equations~\eqref{em-fluid} and~\eqref{em-final} it follows that:

\begin{equation}
  \left( e + p \right) u_\alpha u_\beta - p g_{\alpha\beta} = 
\left(\pm \rho \frac{ \mathrm{d} \lmatt  }{ \mathrm{d} \rho }
    \mp \lmatt \right) g_{\alpha\beta} \mp \rho u_\alpha u_\beta  \frac{
      \mathrm{d} \lmatt  }{ \mathrm{d} \rho }
\label{diff-equation}  
\end{equation}

  In order to find a differential equation for the matter Lagrangian \(
\lmatt \), we proceed in the following four alternative ways:

\begin{enumerate}[(i)]
  \item Equate terms with \( u_\alpha u_\beta \) and the ones with \(
        g_{\alpha\beta} \) in equation~\eqref{diff-equation}, to obtain:
	\begin{gather}
           e + p =  \mp  \rho \frac{ \mathrm{d} \lmatt  }{ \mathrm{d} \rho },
	  					\label{d01} \\
	  - p = \pm \rho \frac{ \mathrm{d} \lmatt  }{ \mathrm{d} \rho }
	          \mp \lmatt.			\label{d02} 
	\end{gather}
  \item Multiply equation~\eqref{diff-equation} by the metric tensor \(
        g^{\alpha\beta} \) to obtain:
	\begin{gather}
	  e - 3p = \pm 3 \rho \frac{ \mathrm{d} \lmatt  }{ \mathrm{d} \rho } 
	          \mp 4 \lmatt.
        \label{d03}
	\end{gather}
  \item Multiply equation~\eqref{diff-equation} by the four velocity \(
        u^\alpha  \) to obtain:
	\begin{equation}
	  \lmatt = \mp e.
	\label{d04}
	\end{equation}
  \item Multiply equation~\eqref{diff-equation} by the projection tensor
        \( P^{\alpha}{ }_\lambda := u^\alpha u_\lambda - 
	\delta^\alpha{ }_\lambda \), which is orthogonal to the four 
	velocity \( u_\alpha \), to obtain exactly equation~\eqref{d02}
\end{enumerate}

  The sets of equations~\eqref{d01}-\eqref{d02}, relation~\eqref{d03}
and~\eqref{d04} seem all contradictory at first sight and as mentioned
in the introduction, their study has caused quite a lot of confusion
for at least 65 years, even  for the simple case of dust, for which \(
p= 0 \) and so \( e = \rho c^2 \).  However, note that according to
equation~\eqref{d04}, \( \lmatt = \mp e \) and the same result is also
obtained by the system of equations~\eqref{d01}-\eqref{d02}.  With the
value of \( \lmatt = \mp e \) equations~\eqref{d01}-\eqref{d03} become:

\begin{equation}
 \rho \frac{ \mathrm{d}  e }{ \mathrm{d} \rho } = e + p,
\label{first-law_01}
\end{equation}

\noindent which is exactly the differential equation satisfied by the first
law of thermodynamics for an ideal fluid as expressed in 
equation~\eqref{first-law-ideal}.

\section{Is the Matter Lagrangian of an ideal flow \( \lmatt = \pm p \)?}
\label{matter-lagrangian-p}

  As mentioned in the introduction, there are many works in which the
matter Lagrangian \( \lmatt = \pm p \).  In general terms the analysis is
based on the following procedure~\citep[see e.g.][and references
therein]{avelino18,ferreira20}, which uses auxiliary scalar fields to find
the value for the matter Lagrangian of an ideal fluid.

  Assume that the matter Lagrangian

\begin{equation}
  \lmatt = \lmatt(\zeta), \quad \text{where} 
    \quad \zeta := \frac{ 1 }{ 2 } | \boldsymbol{\nabla} \phi  |^2 = \frac{1}{2}
    \nabla^\alpha \phi \nabla_\alpha \phi,
\label{a01}
\end{equation}

\noindent for a real scalar field \( \phi \).  Since \( \zeta =g^{\mu\nu}
\nabla_\mu \phi \nabla_\nu \phi / 2 \) and using
relations~\eqref{foot02}-\eqref{foot03} in equation~\eqref{em-co} it
follows that:

\begin{equation}
  T_{\alpha\beta} = \pm \frac{ \partial{ \lmatt } }{ \partial \zeta }
  \nabla_\alpha \phi \nabla_\beta \phi \mp \lmatt g_{\alpha \beta}.
\label{a02}
\end{equation}

  Direct comparison of the previous equation with the value for the
energy-momentum tensor for an ideal fluid given in
equation~\eqref{em-fluid} means that:

\begin{equation}
  u_\alpha := \frac{ \nabla_\alpha \phi }{ \sqrt{2 \zeta } }, \quad 
   e + p = \pm 2 \zeta \frac{ \partial \lmatt }{ \partial \zeta}, \quad 
   \lmatt = \pm p.
\label{a03}
\end{equation}

  Note that the four velocity \( u_\alpha \) defined above is normalised
such that \( u_\alpha u^\alpha = 1 \). This procedure is to be taken
with care since the four velocity needs to be time-like or else the
perfect fluid approximation breaks down. This condition cannot be fully
guaranteed if \( u_\alpha \propto \nabla_\alpha \phi \).  Also note
that using the first law of thermodynamics~\eqref{first-law-ideal} for
a perfect fluid, which by definition is adiabatic, and the fact that \(
\lmatt = \pm p \), the second differential equation in the previous set
of equations can be manipulated so that:

\begin{equation}
  \frac{ \partial{ \ln \zeta^{1/2} } }{ \partial p  } = \frac{ \partial
  \ln \rho }{ \partial e }, 
\label{a04}
\end{equation}

\noindent i.e.:

\begin{equation}
  \ln \zeta^{1/2} \propto \int{ \frac{ \partial \ln \rho }{ \partial
  e } \mathrm{d} p }.
\label{a04b}
\end{equation}

  Since \( \zeta = |\boldsymbol{\nabla} \phi|^2 / 2 \), then the scalar
field \( \phi \) is a purely thermodynamical function and the same occurs
for the four velocity \( u^\alpha \).  In other words, the previous
relation constitutes a general ``constraint'' for the four velocity of any
ideal fluid in any space-time regardless of the constitutive field
equations and so, proposition~\eqref{a01} must be incorrect.  Any proposed
Lagrangian as a function of a scalar field for which its derivatives
represent a four velocity will also be incoherent.  The beauty of the
combined results of equations~\eqref{d01}-\eqref{d03} with \( \lmatt =
\mp e \) is the purely thermodynamic relation~\eqref{first-law_01}, which
is the first law of thermodynamics~\eqref{first-law-ideal}, and is directly
obtained from the metric definition of the energy-momentum
tensor~\eqref{energy-momentum} with no 
auxiliary fields needed to find the correct matter Lagrangian for
an ideal fluid.

\section{Final remarks}
\label{remarks}

  In this work we have shown that the value of the matter Lagrangian 

\begin{equation}
  \lmatt = \mp e,
\label{matter-lagrangian-final}
\end{equation}

\noindent is valid for an ideal fluid with an energy-momentum tensor
given by \( T_{\alpha\beta} = \left( e + p \right) u_\alpha u_\beta -
p g_{\alpha\beta} \).  As mentioned in Section~\ref{matter-lagrangian-p},
other obtained values in the literature that use auxiliary scalar fields
yield inconsistencies since the hydrodynamical properties of the fluid
would be directly related to the auxiliary scalar field, regardless of
the constituent field equations.  The cases for which \( \lmatt = \pm p \)
show a non-conservation of the matter current \( \rho u^\alpha \) meaning
a non-conservation of mass~\citep[see e.g.][]{schutz70,minazzoli12}.

 In the present work we showed that the general result of the matter
Lagrangian for an ideal fluid in equation~\eqref{matter-lagrangian-final}
is in perfect agreement with the conservation of the matter current
through the continuity equation~\eqref{continuity} and coherent with
the fact that an ideal fluid moves adiabatically, according to the first
law of thermodynamics~\eqref{first-law-ideal}.

  For the case of dust, i.e. \( p= 0 \), it follows that: 

\begin{equation}
 e = \rho c^2, \quad \text{and} \quad \lmatt = \mp \rho c^2.  
\label{dust}
\end{equation}

  The total energy density of a barotropic fluid, one for which the pressure
\( p(\rho) \) is a function of the mass density \( \rho \) only, is given
by~\citep[see e.g.][]{landau-fluids}:

\begin{equation}
  e = {\rho} c^2 + \rho \int{ p(\rho) \mathrm{d} \rho / \rho^2  }, 
\label{barotropic-energy}
\end{equation}

\noindent and so:

\begin{equation}
\lmatt = \mp {\rho} c^2 \mp \rho
\int{ p(\rho) \mathrm{d} \rho / \rho^2  },
\label{barotropic}
\end{equation}

\noindent  which coincides with the
results obtained by~\citet{minazzoli12}.  For the case of a polytropic
fluid for which

\begin{equation}
  p \propto \rho^\kappa,
\label{polytropic}
\end{equation}

\noindent with constant polytropic index \( \kappa \), it follows 
that~\citep{tooper}

\begin{equation}
  e = \rho c^2 + \frac{ p }{ \kappa - 1 },
\label{tooper}
\end{equation}

\noindent and so:

\begin{equation}
  \lmatt =  \mp \rho c^2 \mp \frac{ p }{ \kappa - 1 }.
\label{matter-lagrangian-polytrope}
\end{equation}

  The energy density equation~\eqref{tooper} becomes the Bondi-Wheeler 
equation of state~\citep{tooper} for
the case in which the pressure of the fluid is much greater than its rest
energy density \( \rho c^2 \), i.e. \( p \gg \rho c^2 \) and so:

\begin{equation}
  e = \frac{ p }{ \left(\kappa -1 \right) }.
\label{bondi-wheeler-state}
\end{equation}

\noindent For this Bondi-Wheeler case the matter Lagrangian is given
by:

\begin{equation}
  \lmatt = \mp \frac{ p }{ \kappa - 1 }.
\label{bondi-wheeler}
\end{equation}

  In  cosmological applications the previous two relations
are valid for the cases of radiation with \( \kappa = 4/3 \), a monoatomic
gas with \( \kappa = 5/3 \) and for cosmological vacuum with \( \kappa =
0 \).  The case of cosmological dust is represented by a matter Lagrangian
given by equation~\eqref{dust}.\footnote{For the choice of a
(\(-,+,+,+\)) signature, since the right-hand side of 
equation~\eqref{four-velocity} equals \( -1 \), and the energy-momentum
tensor in equation~\eqref{em-fluid} is given by:
\begin{displaymath}
  T_{\alpha\beta} = \left( e + p \right) u_\alpha u_\beta + p
  g_{\alpha\beta},
\end{displaymath}
\noindent it follows that equation~\eqref{deltarho} turns into:
\begin{displaymath}
  \delta \rho = \frac{ 1 }{ 2 } \rho \left( g_{\alpha\beta} + u_\alpha
    u_\beta \right) \delta g^{\alpha\beta}.
\end{displaymath}
\noindent With all this, the \( \pm \) and \( \mp \) signs that appear in
equations~\eqref{d01}-\eqref{d03} are inverted so that:
\begin{equation}
  \lmatt = \pm e,
\end{equation}
\noindent which leaves equation~\eqref{first-law_01} unchanged.  This means
that for this metric signature, the \( \pm \) and \( \mp \) signs on
equations~\eqref{dust}-\eqref{barotropic}
and~\eqref{matter-lagrangian-polytrope}-\eqref{bondi-wheeler} are also
inverted.
}

\section*{Acknowledgements} 
This work was supported by DGAPA-UNAM (IN112019) and Consejo Nacional
de Ciencia y Teconolog\'ia (CONACyT), M\'exico (CB-2014-01 No.~240512)
grants. SM acknowledges economic support from CONACyT (26344).

\bibliographystyle{apsrev4-2}
\bibliography{mendoza-silva}

\end{document}